# Nano-Ratchet sets free electrical power even without intentional excitation

H. Weidlich[1*], V. Hortelano[1,2], W. T. Masselink[2], G. Mahler[3], R. Hey[4] and Y. Takagaki[4]

[1] Institute Prof. Dr. Georg Kurz GmbH, Stöckheimer Weg 1, 50829 Cologne, Germany
[2] Department of Physics, Humboldt University Berlin, Newton Str. 15, 12489 Berlin, Germany
[3] Institute of Theoretical Physics, Stuttgart University, Pfaffenwaldring 57//IV, 70550 Stuttgart, Germany
[4] Paul Drude Institute for Solid State Electronics, Hausvogteiplatz 5-7, 10117 Berlin, Germany
e-mail:[*]h.weidlich@institut-kurz.de

**In theory ratchets enable energy harvesting by rectifying electrical fluctuations in conductors and semiconductors. Certain rectification effects in nanostructures have been demonstrated earlier. But these were not found to be usable for energy harvesting, as they were not accomplished by asymmetric transmission probabilities in the nanochannels, but rather by other secondary effects. We propose and experimentally prove a nano-ratchet effect directly linked to an asymmetry in the equilibrium transmission probability, which becomes possible when a magnetic field is present. Remarkably, the current/voltage do not disappear even without any intentional external excitation. The rectification is manifested to be highly efficient to the extent that generation of energy out of environmental electrical noises is achieved.**

Nano-ratchets in theory are one of the technologies to extract usable energy out of random fluctuations. Ratchets have been discussed intensively in theoretical physics[1,2]. Devices designed to rectify randomly moving electrons in metals or semiconductors have been proposed. They have been called "electrical ratchets"[3,4]. Electrical ratchet effects have been reported using various types of nano-structures, including asymmetrically-shaped channels[5], asymmetrically-shaped scatterers embedded in channels[6,7], double quantum dot configurations[8] and superconductor-based circuits[9]. However, for all two-terminal devices Onsager's symmetry applies: $T_{ij}(B) = T_{ji}(-B)$, where $T_{ij}(B)$ is the probability of transmission from lead $j$ to lead $i$ in the presence of a magnetic field $B$[9]. This implies that in the absence of a magnetic field the transmission probability is always symmetric regardless of the channel geometry and structure (more references see Supplementary Section 1).

We propose in this paper a new mechanism for the ratchet effect in a two-terminal configuration. A rectification effect is induced in the presence of a static magnetic field, which was not the case in the previous reports. The transmission probability is thereby allowed to be asymmetric. Our work is the first ever demonstration of the ratchet effect in two-terminal channels based on an asymmetry that is realized even on the level of the equilibrium transmission probability. We show that such ratchet effect is robust. Moreover, a finite ratchet current/voltage is observed even in the absence of any intentional excitation. We interpret the current to be generated by rectifying random electrical fluctuations inside the conductor/semiconductor. The observation evidences marked sensitivity of the mechanism and gives a hope that the device can harvest energy from background electrical noises.

To explain the mechanism for a ratchet effect based on an asymmetric transmission probability, let us consider the horse-shoe-shaped narrow channel illustrated in Fig. 1a. The curved geometry makes it imminent that the electrons are multiply reflected from the channel boundary in the ballistic transport regime. In channels having a uniform width $W$, the incident electrons are always transmitted in the forward direction of the channel in the classical situation when the reflection from the boundary is specular. The electrons can be backscattered if the boundary reflection is, at least partly, diffuse as they are reflected into arbitrary directions. The probability of transmission in the latter circumstance becomes asymmetric when ballistic trajectories in the clockwise and counterclockwise directions are altered in different ways by bending the electron motion with the Lorentz force in the presence of a magnetic field. On the one hand, the number of the boundary reflections decreases if the Lorentz force turns the trajectory in the direction of the curving of the channel. The number is expected to be minimal when the cyclotron radius $r_c$ is comparable with the curvature radius $R$ of the channel, see the trajectory α in Fig. 1a. The boundary reflection, on the other hand, becomes frequent when the electron trajectory and the channel bend in the opposite directions, as shown by the trajectory β in Fig. 1a. That is, the diffuseness of the boundary scattering results in an asymmetry in the transmission between the electrons propagating from the left-hand-side end to the right-hand-side end and those propagating in the opposite direction as the average number of the boundary reflections differs between them.

From the viewpoint of general transport theory, the system behaves ergodic as long as the channel boundary scatters the electrons specularly. Following the theorem of Liouville, this holds true not only for the specular reflection but also for a reflection where the collision with the wall occurs with the sinus function probability, i.e., the probability of outgoing is the highest for the direction perpendicular to the wall and the lowest for the direction parallel to the wall[10]. The system behaves as a non-ergodic open system, however, as soon as the presence of other types of reflections from the wall is assumed[11].

We have evaluated the transmission probabilities between the two ends of the channel using the billiard model[12]. The reflection from the channel boundary was assumed to be only partly specular. The electrons were reflected from the wall preserving the incidence angle with probability $p$ but were scattered in random directions with probability $1-p$. In Figs. 1b and 1c, we show the prevalence for the preferred side defined as $P = T_{cw} - T_{ccw}$, where $T_{cw}$ and $T_{ccw}$ are the transmission probabilities in the clockwise and counterclockwise directions, respectively. As expected, a peak occurs in $P$ at $R/r_c \approx 1$, as shown in Fig. 1b. The transmission asymmetry increases as the boundary reflection is made diffuse, as one finds for the filled circles in Fig. 1c. (see Supplementary Section 2)

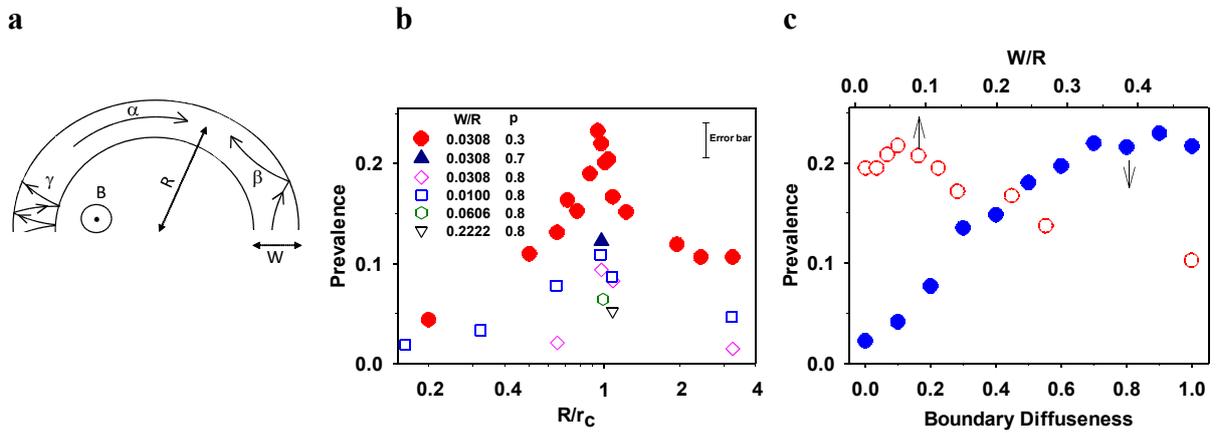

**Figure 1 | Theoretical simulations of ratchet effect. a,** Cyclotron orbits in a horse-shoe-shaped narrow channel. A magnetic field $B$ is applied perpendicular to the plane. The ballistic trajectories α-γ show typical orbits relevant for the ratchet effect. The curvature radius and width of the channel are $R$ and $W$, respectively. **b,c,** Prevalence when $R/r_c$, boundary diffuseness $1-p$ or $W/R$ is varied. The prevalence $P$ is defined as the difference between the transmission probabilities in the clockwise and counterclockwise directions of the horse-shoe channel. The parameter $R/r_c$ with $r_c$ being the cyclotron radius is proportional to $B$. $l_e/R \approx 1.23$ was assumed. In **c,** $r_c/R = 1.026$, $W/R = 0.0307$ and $l_e/R = 1.23$ for the filled circles and $p = 0.3$ for the open circles. For the open symbols in **b,** the angular distribution for the electron injection was assumed to be isotropic. Cosine dependence was assumed in the rest of the cases.

In order to verify experimentally the ratchet mechanism proposed above, we employed nano-channels defined using microfabrication technologies in the two-dimensional electron gas (2DEG) of a selectively doped GaAs-(Al,Ga)As heterostructure (see Supplementary Section 3). Figure 2a shows the scanning electron micrograph of a U-shape structure. The scattering of conduction electrons from the channel boundaries defined using dry etching techniques has been revealed to be partly diffuse[13]. While the reflection from the channel boundary created using electrostatic depletion is almost specular, the ion-beam exposure in dry etching makes the boundary considerably rough[14,15].

In Fig. 2b, we show the current measured between the two-terminals of the device in Fig. 2a. The measurements were carried out at a temperature of 10 K in the presence of a magnetic field $B$. For the filled symbols, the magnetic field was applied perpendicular to the heterostructure. While the current at $B = 0$ was absent if we ignore the offset, a current emerged when the magnetic field was strengthened. One finds that the curve is approximately antisymmetric with respect to $B = 0$ and the current maximized at $|B| = 0.2 - 0.3$ T. We attribute this current to the ratchet effect. We have confirmed that the current remained to be almost absent when the magnetic field was applied parallel to the heterostructure, as shown by the open circles in Fig. 2b. The current is thus indicated to have originated from the cyclotron orbital motion of the 2DEG.

As predicted in Fig. 1b, $R/r_c \approx 1$ is expected at the peak of the ratchet current. For the case in Fig. 2b, we obtain $r_c = \hbar k_F/eB = 393$ nm at $B = 0.2$ T. Here, $k_F = (2\pi n_s)^{1/2}$ is the Fermi wavenumber with $n_s$ being the sheet electron concentration. In GaAs-(Al,Ga)As heterostructures, surface depletion from the sidewall of an etched mesa is large (about 200 nm in our samples). The influence of the depletion is particularly significant for L-shaped corners, where the sharp corners are rounded in forming actual conduction geometry. One can expect the conduction channel in the device shown in Fig. 2a to be approximately horse-shoe-shaped, as illustrated by the blue dotted curves. The large depletion makes it difficult to estimate the effective curvature radius from the experimental geometry. Nevertheless, the cyclotron orbit with the radius of 393 nm fits reasonably well with the effective conduction channel, as shown by the red semicircle in Fig. 2a.

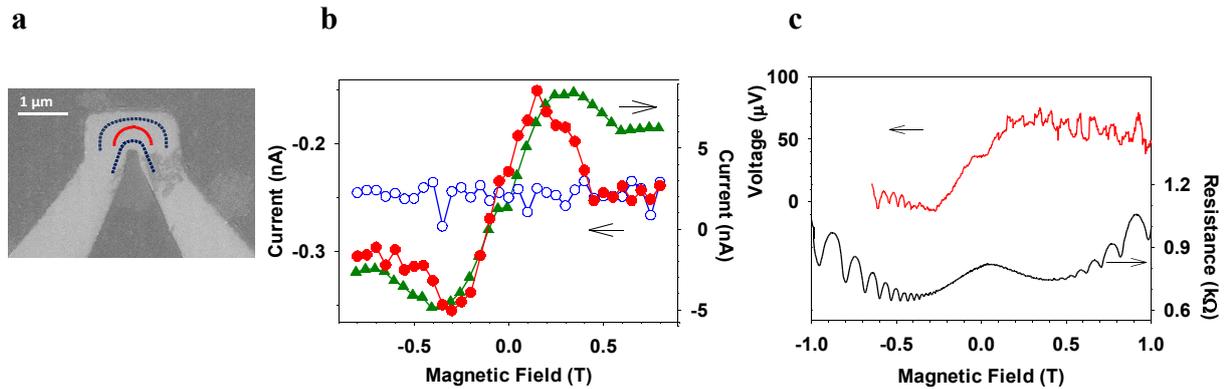

**Figure 2 | Magneto-ratchet effect in U-shaped narrow channel. a,** Scanning electron micrograph of a U-shaped channel in a GaAs-(Al,Ga)As heterostructure. The red semicircle illustrates the cyclotron orbit at the magnetic field (0.2 T) for the peak in the ratchet current in **b**. The blue curves depict narrowing of the effective conduction channel from the geometry due to lateral surface depletion. **b,** Ratchet current induced by magnetic field. The device in **a** was measured at a temperature of 10 K. The direction of the magnetic field with respect to the structure was perpendicular for the filled symbols and parallel for the open circles. Rf excitation with an amplitude of 1 V and a frequency of 50 MHz was applied to the device for the filled triangles. For the circles, no external excitation was given. **c,** Comparison of ratchet voltage with resistance. The upper curve shows the open-circuit two-terminal voltage without external rf excitation. The four-terminal longitudinal resistance obtained using the lock-in technique is shown by the lower curve. These measurements were carried out at a temperature of 0.3 K with the magnetic field applied perpendicular to the device.

When the temperature was lowered to 0.3 K, an oscillation emerged as superimposed on the two-terminal voltage, as shown by the upper curve in Fig. 2c. The oscillation was practically identical with the Shubnikov-de Haas oscillation in the four-terminal resistance of the channel shown as the lower curve in Fig. 2c, providing additional evidence that the 2DEG was responsible for the finite current/voltage that emerged under the application of the magnetic field. The resistance in Fig. 2c exhibited a broad peak around zero magnetic field. The peak was presumably produced by the confinement in the horse-shoe channel[16]. The peak amplitude is expected to be proportional to the difference in the number of conducting states in the narrow channel and in the wide 2DEG areas attached to the channel[17]. The peak thus disappears when $2r_c < W$. This effect is absent for the voltage in Fig. 2c, suggesting that the two-terminal current/voltage is almost completely dominated by the electrons in the narrow channel without being influenced by the surrounding 2DEG (see Supplementary Section 4).

The presence of a remaining current even in case of absence of external excitation is shown in Fig. 3 for cases with magnetic fields applied to realize the ratchet effect and of $B = 0$. The current remained to be almost zero when $B = 0$ but increased nearly antisymmetrically with an external excitation voltage $V_{rf}$ when the magnetic field was optimum for the ratchet effect. We emphasize that no indication of sign change was noticed along all the way of $V_{rf}$-increase opposite to what was as often encountered by previously proposed nano-ratchets, (see Supplementary Section 1) and the roughly linear increase of the current with the excitation did not break down even for the strongest excitation. As we show by the filled triangles in Fig. 2b, the antisymmetric shape hardly changed at $V_{rf} = 1$ V, manifesting the robustness of the ratchet effect. Remarkably, the ratchet current did not disappear when the external excitation was reduced to zero, see the inset of Fig. 3. The data plotted by the filled circles in Fig. 2b were taken, in fact, in the absence of the external excitation. Thus the current at zero excitation was generated by rectifying the random electrical noises in the background. We have carried out these experiments using electronic multimeters (which might be susceptible to stray-voltages by their own inner design) as well as by using high sensitivity analogue galvanometers (which works with internal battery). The ratchet effect in our devices is evidenced to be highly efficient and capable of harvesting energy out of environmental random fluctuations[18].

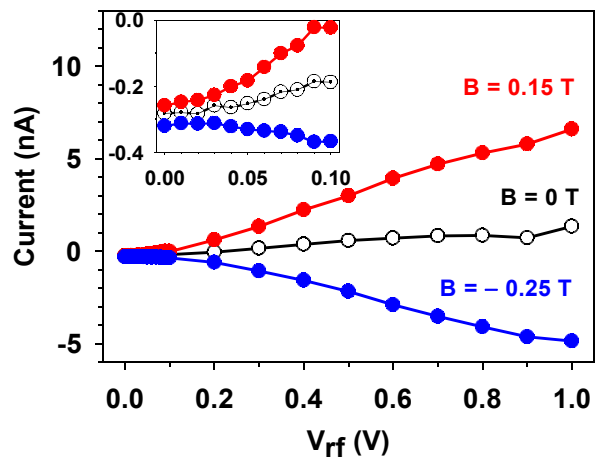

**Figure 3 | Dependence of ratchet current on amplitude $V_{rf}$ of external rf excitation.** The current was measured at magnetic fields of $B = -0.25$, 0 and 0.15 T at a temperature of 10 K. The ratchet current became maximum at the magnetic fields of $-0.25$ and 0.15 T in this measurement run. The frequency of the rf excitation was 50 MHz. The weak excitation part is shown with expanded scales in the inset.

The temperature dependence of the ratchet effect is shown in Fig. 4a. The ratchet current decreased in amplitude when the temperature $T$ was raised from 16 K to 20 K. The antisymmetric shape remained at $T = 20$ K as shown with a magnified scale in Fig. 4a. The antisymmetric characteristic disappeared when the temperature was higher than 25 K. The disappearance is attributed to the reduction of the mobility at high temperatures. In Fig. 4b, the dependence of the prevalence $P$ on $l_e$ was simulated. The prevalence is almost unchanged for large values of $l_e$ and drops rapidly as $l_e$ reduces to zero. It appears that the transition occurs when $l_e$ is comparable to $\pi R$, i.e., the length of the horse-shoe channel.

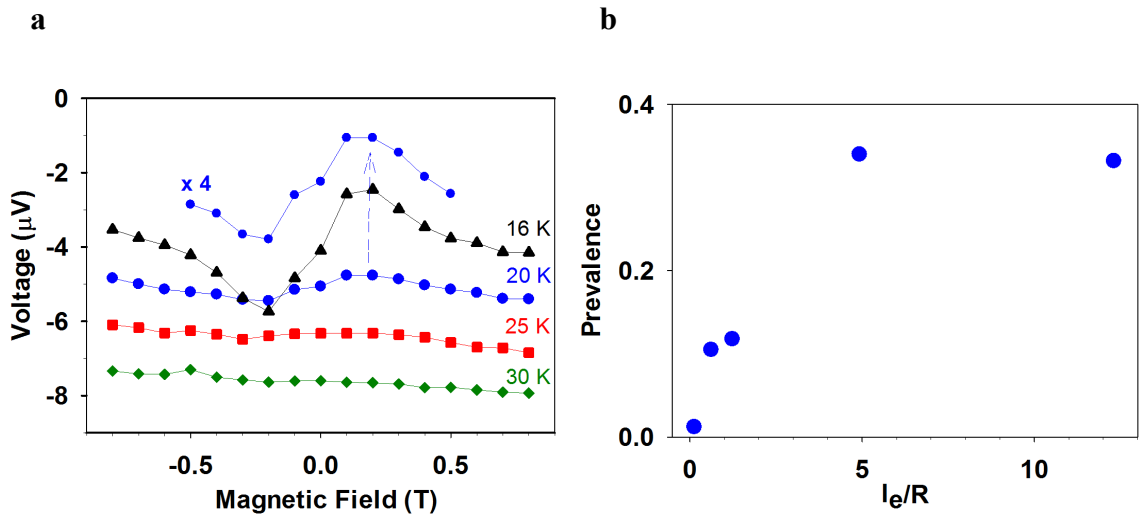

**Figure 4 | Temperature limit of ratchet effect. a,** Temperature dependence of ratchet effect. The two-terminal voltage was measured when the external excitation was absent. The curve at the temperature of 20 K is shown also with a magnification by a factor of 4. **b,** Simulated prevalence in horse-shoe channel when mean free path length $l_e$ is varied. The cyclotron radius is $r_c = 1.026R$ and width of the channel is $W = 0.0307R$, where $R$ is the curvature radius of the channel. The specularity of the boundary is $p = 0.3$.

In summary, we have demonstrated a mechanism that enables a robust ratchet effect in a controllable manner by utilizing the ballistic transport in nanofabricated semiconductor channels. The electron transport in horse-shoe channels becomes asymmetric when the ballistic trajectories are bent by applying a perpendicular magnetic field under the circumstance that the reflection from the channel boundary is not completely specular. Remarkably, a non-zero current is present even without the intentional external excitation. The rectification is manifested to be highly efficient to the extent that extraction of energy out of environmental electrical noises is possible.

**Acknowledges**

The authors thank Alexander Kusdas from Sammlung historische Messtechnik for providing the galvanometer and Julia Dobbert for her collaboration in the initial stage of the experiments.

**Author contributions**

H.W. conceived the idea performed the numerical simulations and organized the whole project. V.H. prepared the samples and carried out the experiments. H.R. provided heterostructures. W.T.M. guided the initial stage of the experiments and supplied measurement facilities. Y.T. guided the experiments of the ratchet effect. Y.T., V.H. and H.W. wrote the manuscript. All authors discussed the results and commented on the manuscript.

**Supplementary information**

### 1. Ratchet effect

The ratchet effect observed in the previous reports was not due to asymmetry in the transmission probability but was associated with secondary asymmetry effects that resulted from non-equilibrium potential landscapes under external biases[1]. The consequential shortcomings become apparent if one considers the fact that the direction of the output current is not obviously predictable as it is typically determined by competing aspects. The operation of the ratchets was not satisfactorily stable and suffered from, for instance, changes in the sign of the current when excitation intensity[3] or temperature was varied[1]. To overcome this fundamental problem associated with the Onsager symmetry, a four-terminal configuration was employed in refs. (1,2) to take advantage that the four-terminal resistance obeys different symmetry relationships, such as it does not have to be identical under magnetic field reversal[4].

The ratchet effect observed in such a situation in the previous reports was not due to asymmetry in the transmission probability but was associated with secondary asymmetry effects that resulted from non-equilibrium potential landscapes under external biases[1]. The consequential shortcomings become apparent if one considers the fact that the direction of the output current is not obviously predictable as it is typically determined by competing aspects. The operation of the ratchets was not satisfactorily stable and suffered from, for instance, changes in the sign of the current when excitation intensity[3] or temperature was varied[1].

## 2. Simulation

When the mean free path $l_e$ of an electron is much longer than its deBroglie wavelength, the movement of the electron can be described by classical mechanics without invoking quantum mechanics. The motion of the electrons in the horse-shoe channel can be therefore simulated by a billiard-type model[5]. A large number of electrons were injected from one end of the channel and their trajectories were followed until they reach one of the ends of the channel. The numbers of the electrons either transmitted to the other end or reflected back to the incident end provide the transmission and reflection probabilities, respectively. On injecting the electrons, the spatial distribution was assumed to be uniform over the channel width. The angular distribution was taken to be $P(\theta) = \cos(\theta)/2$, where $\theta$ is the angle of incidence with respect to the channel axis and is distributed in the interval $[-\pi/2, \pi/2]$.

Hard-wall confinement in the channels was assumed in the simulations, and so the electrons were injected with the cosine dependence for the angular distribution[5]. The angular distribution remains to be the cosine dependence even when the confinement potential in the channels is, for instance, parabolic[6]. Nevertheless, if the angular distribution is assumed to be isotropic, $P$ is obtained to be 0.084, whereas $P = 0.069$ for the cosine distribution when $W/R = 0.0307$ and $p = 0.8$. (Note that the results in Fig. 1b include cases where isotropic injection was assumed, which is distinguished using open symbols.) The electrons injected at angles nearly perpendicular to the channel axis undergo a large number of boundary reflections as they bounce between the two sides of the channel, see the trajectory $\gamma$ in Fig. 1a. The influence of the boundary scattering is, as a consequence, more significant for the isotropic injection than for the cosine injection. The above example suggests that the bouncing trajectory $\gamma$ contributes more to the asymmetric transmission than the reflectionless ideal trajectory $\alpha$, which is favored by a collimated injection such as the cosine angular distribution. The prevalence is stronger for smaller $W/R$, as shown by the open circles in Fig. 1c. Narrowing the channel enhances the ratchet effect presumably as the influence of the diffuse boundary is made significant, similar to the role of the trajectory $\gamma$.

With respect to the dependence ratchet-effect on the assumed mean-free-path-length (as shown in Fig. 4b it is note-worthy that the impurity scattering potential in GaAs-(Al,Ga)As heterostructures is known to be long-ranged. The elastic scattering from the impurities is, as a consequence, anisotropic favoring forward scattering. The elastic scattering in the calculations was, however, assumed to be isotropic. The anisotropy in the impurity scattering gives rise to $l_e$ determined by the mobility overestimating the ballistic length that characterizes the decay of ballistic transport phenomena[7]. We point out that the forward scattering plausibly plays a negligible role for the ratchet effect as such small-angle scatterings barely amount to switching the exiting lead of the horse-shoe channel. In other words, electrons are guided to maintain the propagating direction by the channel boundary until they are scattered backwards by a large angle scattering.

## 3. Experimental methods

A selectively doped GaAs-(Al,Ga)As heterostructure was grown by molecular beam epitaxy. The sheet electron density and the electron mobility at a temperature of 0.3 K were $n_s = 2.3 \times 10^{15}$ m$^{-2}$ and 180 m$^2$/Vs, respectively. The elastic mean free path is estimated to be $l_e = 14$ μm. U-shaped narrow channels were defined in the heterostructure using the electron-beam lithography and dry etching techniques. The Ar ion milling was carried out with an acceleration voltage of 0.6 keV. The ohmic contacts were prepared by depositing Ni (5 nm), Au$_{0.88}$Ge$_{0.12}$ (100 nm), Ni (25 nm) and Au (75 nm) layers as a stack. The inclusion of the bottom Ni layer on the GaAs surface widens the optimum temperature range for alloying the ohmic contact[8]. This allowed us to perform the alloying at a high temperature to anneal the damages induced in the dry etching process.

Unless stated otherwise, electrical measurements were carried out at a temperature of 10 K with a magnetic field applied perpendicular to the heterostructure. The frequency of the rf excitation was 50 MHz. The measurements of the current and voltage were perform using a 419 DC Null Voltmeter (Hewlett Packard), which operates using an internal battery. For measuring the current under no external excitation, a mirror galvanometer (Siemens & Halske) was also employed to compare the results.

## 4. Experimental results

The antisymmetric behaviours of the ratchet effect in Figs. 2b and 2c are, as a matter of fact, surprising. One may describe the current induced in response to an external bias in terms of the resistance of the device. Contrary to the four-terminal resistance, the two-terminal resistance is required to be symmetric under magnetic field reversal, i.e., $R_{2t}(B) = R_{2t}(-B)$. The antisymmetric shape of the characteristics, therefore, implies that the quantity that we measured was the transmission probability rather than the resistance. Violations of the Onsager symmetry have been explored in nonlinear transport regime[9-11] or when electron-electron interaction plays a role[12]. In view of the linearity, the resistance of a ratchet is inherently nonlinear. Rectification involves changes in not only the sign but also the magnitude of the current when the direction of the external electric field is reversed. The non-identical slopes in the current-voltage characteristics under the voltage reversal can be expressed by including the higher order voltage terms in the manner of the Fourier series. The magnetic field reversal symmetry of the two-terminal resistance, which should be satisfied for the linear transport coefficient, is not relevant[10]. For AC excitations, the linear order current of a ratchet is on average zero $<i_V> = 0$. The second order current provides a non-zero contribution $<i_V^2>$, which is predominantly antisymmetric under magnetic field reversal[13]. For the effective DC excitation in the resistance measurement in the lower curve in Fig. 2c, the linear term dominates, giving rise to the symmetric behavior.

The ratchet effect in our devices is evidenced to be highly efficient and capable of harvesting energy out of environmental random fluctuations[14.] As to the cause of such fluctuations (fluctuations induced by stray-currents of the electrical magnet used to generate the static electrical field, induced by the electrically driven cryostatic device, induced by stray-electromagnetic fields in the experimental room or the pure thermodynamic statistically random fluctuations of conduction electrons in rigid bodies) it will be the subject of our future series of experiments to conduct these measurements using permanent magnets (instead of electrical magnets), Faraday-shielding of the whole set up, Dewar-technology to generate the cryostatic conditions without use of electrical devices and again analogue galvanometer in order to switch off one possible source for excitation after the other, until the pure statistical random fluctuations inside the experimental sample are left.